# Theoretical analysis of the performance of a diffraction grating back-reflector in infrared-sensitive solar cells

Mario M. Jakas and Francisco Llopis

*Departamento de Física Fundamental y Experimental, Electrónica y Sistemas*

*Universidad de La Laguna, 38205 La Laguna, Tenerife Spain*

**Abstract**

The increase of cell efficiency resulting from using a diffraction grating as a back reflector is investigated. An enhancement coefficient is introduced as a figure of merit that accounts for the ability of the rear grating to increase the generation of electric carriers. According to results in this paper, a diffraction-grating-like relief on the rear face of the photovoltaic cell may reduce the intensity of the specular reflection while favouring the intensity of high-angle rays. In this way, the path-length of reflected rays increases, and so does the efficiency of the cell.

## 1. Introduction

While approximately thirty percent of the energy in the solar spectrum lies between 1 and 2.5 μm [1], the spectral sensitivity of typical silicon or AsGa photovoltaic (PV) cells becomes negligible beyond 1 μm. It is obvious, therefore, that by extending the spectral sensitivity to infrared (IR) range the efficiency of the PV cell could be significantly increased [2, 3]. However, extending the sensitivity of PV cells to IR without reducing the efficiency in the visible range of the spectrum constitutes a major challenge. Whatever be the solution to this problem one thing is for sure: since most semiconductor materials have a relatively low absorption coefficient in the IR range [4], no such high-efficiency cells can be conceivably devised without trapping the IR-light as well.

Several approaches have been proposed to confine light within the PV cells [5]. Texturing the top surface is possibly the most widely used way of trapping light. Back-reflectors have proven to be feasible too. In this regard, Campbell et al. [9] devised a pyramidally textured rear surface which nicely acts as a Lambertian reflector. Heine et al. [6] analysed the possibility of using a diffraction-grating-like relief coated with Ag and, more recently, Zeng et al. [7] built a diffraction grating reflector where the Ag coating is replaced by a distributed Bragg reflector (DBR). All these approaches seem to improve the external efficiency of the thin cells. Unfortunately, however, such an improvement decreases as the thickness of cell increases [6, 7].

It must be borne in mind though, that in the case of IR the difficulty above may not necessarily apply. As a consequence, none of the previously mentioned approaches should be ruled out. Actually, according to the results in this paper, a plain diffraction grating, i.e. without Ag-coating and no DBR structure, appears to do a good job. It turns out that by properly choosing the grating shape and size, the intensity of small-angle rays can be reduced while increasing that of more tilted ones. In this way, weakly absorbed light, such as IR, may have a larger path-length within the active layer of the cell, thus augmenting the probability of light for producing electric carriers. Although the present study is to some extent similar to those in Refs.[6, 7], it must be emphasized that they calculated the reflection coefficient rather than the energy deposited by the reflected rays within the active layer. These two quantities are equivalent for visible light, but this does not hold for weakly absorbed light such as IR. In the latter case, the path-length and so the propagation angle of light, is a relevant parameter too. Finally, it is worth mentioning that there seems to be no need for using the DBR or the Ag-coating, because the total reflection in the semiconductor-vacuum interface appears to work fairly well [8]. Furthermore, it is found that a diffraction grating may increase the path-length of rays even more than a Lambertian back-reflector.

## 2. Theory

Suppose a thick PV cell with a diffraction grating texture on its rear face as sketched in figure 1. Both the front and rear faces are assumed to be perpendicular to the *z*-axis, and the diffraction grooves are parallel to the *y*-axis. Somewhere below the surface lies the active layer, which is the region where the electron-holes generated by the incoming light may be used to produce electricity. Assuming that a beam of monochromatic light arrives to the rear face from within, it scatters of the diffraction grating and splits into several reflected rays. The angles of these rays obey the equation

$$\sin\theta_n = \sin\theta_i + \frac{\lambda n}{D}, \qquad (1)$$

where $\theta_i$ is the angle of incidence, $\theta_n$ denotes the angle of the $n$-th diffracted order, $\lambda$ is the wavelength of light within the cell and $D$ is the groove spacing.

The rate of electron-hole pairs produced by the reflected rays per unit area perpendicular to the $z$-axis can be readily calculated as:

$$G_r = \frac{1}{\lambda_{eh}} \sum_n I_n\, p_n = \frac{W}{\lambda_{eh}} \sum_n \frac{I_n}{\cos\theta_n}, \qquad (2)$$

where $I_n$ is the intensity of the $n$-th diffraction order, i.e. energy per unit time and area perpendicular to the $z$-axis, $p_n$ is the path-length of such a beam along the active layer and $\lambda_{eh}$ is the mean-free-path of the light for producing one electron-hole pair.

Dividing Eq. (2) by $W I_i / \lambda_{eh}$ one obtains a dimensionless figure that can be referred to as the *enhancement coefficient*

$$\eta_{DG} = \sum_n \frac{I_n}{I_i \cos\theta_n}. \qquad (3)$$

This expression quantifies the excess of carriers introduced by the diffraction grating. Alternatively, one may use the *diffraction efficiency* of the $n$-th order, namely $\eta_n = I_n/I_i$. Therefore, Eq.(3) can be rewritten as

$$\eta_{DG} = \sum_n \frac{\eta_n}{\cos\theta_n}. \tag{4}$$

This equation clearly shows that the efficiency of the cell can be increased not only by increasing the reflectivity, but also by producing large-angle rays, i.e. *cos θ$_n$* ≈ 0.

In the following section, the results of calculating Eq. (4) using the GD-Calc code [10] and the so-called C-method [11] are presented. For a comparison, ray-tracing calculations are also performed. To this end, a computer code which utilises standard ray-tracing techniques is developed. In all cases, a non-symmetric sawtooth diffraction grating is assumed and, as indicated in figure 1, the grooves have height *h* and width of the left-hand side of the tooth *d*. Similarly, a homogeneous non-absorbing media with a refraction index $n_1$ = 3.6 is assumed and the presence of a top face is ignored.

In connection with this, it must be noticed that all reference to IR light in the present calculations is contained in the assumed value of the refractive index and in the fact that absorption coefficient is neglected. Moreover, the front face, though important, is purposely disregarded here, because its presence will introduce such a large number of free parameters that, unless one limits oneself to a particular case, the extent of this paper would be substantially larger.

## 3. Results and discussions

*3.1. Ray tracing*

Since ray tracing simulations are faster than wave-optics calculations, the former is firstly used in order to produce a coarse sweep over the various parameters entering this model. The results of such calculations are plotted in figure 2 where $\eta_{DG}$ appears as a function of the aspect ratio $h/D$ and a set of $d/D$ ratios, ranging from 0.1 to 0.5. The incident ray is assumed to be TM-polarized and arriving from above at normal incidence, i.e. $\theta_i = 0$.

In the first place, one observes that the enhancement coefficient is a rather oscillating function of $h/D$. This is certainly not caused by statistical fluctuations, since the relative uncertainty of the results in figure 2 is of the order of, or less than one percent. However, a more detailed analysis of the beam trajectories shows that these peaks are produced by the occurrence of certain paths which, through multiple reflections, lead to large reflection angles. Furthermore, it can be observed that, depending on $d/D$, $\eta_{DG}$ has a neat absolute maximum around $h/D \cong 0.2 - 0.8$.

From these results, it turns out that the largest maximum for the cases studied here, belongs to the symmetric sawtooth, i.e. $d/D = 0.5$, with an aspect ratio close to 0.5. For this case the enhancement coefficient becomes approximately equal to 2.8, which is larger than that of a Lambertian reflector with unit reflectivity.

Cases where the incident light is uniformly distributed over a cone are also investigated. The axis of the cone is parallel to the vertical direction and different opening angles are considered. The enhancement coefficients so calculated, however, are not shown here since they are all very similar to the previously discussed results. As a matter of fact, even an opening angle as large as 45 degree produces enhancement coefficients that scarcely differ from those of a well collimated, vertically incident beam.

Similarly, it must be mentioned that identical calculations as those in Fig.2 are carried out for TE polarized light, too. The results, again, can hardly be distinguished from the TM cases and are not worth plotting. However, the fact that TE- and TM-polarized light produce nearly the same results is not a surprise, since most rays arrive at the bottom surface at angles larger than the critical angle and so, they are totally reflected irrespective of the polarization state.

*3.2. Wave-optics calculations*

Having found the optimal size and shape of the diffraction grating, the GD-Calc code [10] is used to calculate the enhancement coefficient. Figure 3 shows the result of such calculations for normal incidence and a symmetric grating, i.e. $d/D = 0.5$. As one can readily see the enhancement coefficient is an oscillatory function of $\lambda/D$, becoming very large when $\lambda/D = 1/n$, $n$ being an integer number. The rightmost peak, for example, corresponds to $n = 7$, i.e. $\lambda/D = 0.143$. As $\lambda/D$ decreases, however, the amplitude of the oscillations decreases approaching a limiting value which achieves the maximum for the $h/D = 0.4$ case. In an average sense, however, the enhancement coefficient does not vary with increasing the wavelength. In order to avoid a busy plot, the results for $h/D = 0.3$ and $0.2$ are plotted in figure 4. The trend observed for the limiting values is in agreement with previous ray-tracing calculations. Comparing with the results in figure 3 it is evident that, as the relief goes shallower, $\eta_{DG}$ decreases and becomes less oscillatory as well. Actually, the number of peaks appears to be smaller.

For the sake of double checking, the enhancement co efficient is also calculated using the so-called C-method [11]. The results, which are not shown here, are observed to compare remarkably well with GD-Calc calculations.

## 4. Conclusions

The advantage of using a diffraction grating as a back-reflector in thick photovoltaic cells is investigated. To this end, ray-tracing and wave-optics methods are used to calculate the enhancement coefficient, namely the excess of electric carriers introduced by the diffraction grating relative to the case of having no back-reflector. According to the present results, by choosing the size and shape of the diffraction grating, the enhancement coefficient can be increased nearly two times that of a cell without a back reflector and 1.5 times that of an ideal Lambertian reflector. This seems to be a consequence of the large reflection angle that reflected rays may have after being diffracted, thus increasing the path-length of these rays within the active layer of the cell. Naturally, these conclusions apply to infrared light, for which both Si and GaAs exhibit low absorption coefficient and a good confinement is therefore advisable. It is worth mentioning that there seems to be no need for using either a metallic coating or a distributed Bragg reflector (DBR) below the diffraction grating as proposed in [6, 7], since the internal total reflection of the semiconductor-vacuum interface suffices.

**Acknowledgements**

The authors would like to thank I. Tobías, A. Luque and A. Martí for interesting discussions concerning the behaviour of back reflectors. This work has been supported in part by the European Union 6th Framework Program FULLSPECTRUM (SES-CT-2003-502620) and the Spanish Education Ministry through the Plan Nacional Consolider project GENESIS-FV (CSD-2006-0004).

**References**

[1] Reference AM 1.5 Solar Spectra. The American Society for Testing and Materials (ASTM). This data can be downloaded as a plain text file from the web site: http://rredc.nrel.gov/solar/spectra/am1.5/

[2] A. Luque et al., Solar Energy Mater. Sol. Cells 87 (2005) 467.

[3] A. Martí et al., Thin Sol. Films 511-512 (2006) 638.

[4] Forouhi et al., Phys. Rev. B 38 (1988) 1865.

[5] J. Nelson, The Physics of Solar Cells, Imperial College Press, London, 2003.

[6] Heine, C. and Morf, R. H., Appl. Optics 34 (1995) 2476.

[7] L. Zeng et al., Appl.Phys.Lett. 89 (2006) 111111.

[8] Llopis, F. and Jakas, M.M., Poster presented at the 23rd European Photovoltaic Solar Energy Conference, Valencia, Spain (2008).

[9] Campbell, P. and Green, M.A., J. Appl. Phys. 62 (1987) 243.

[10] GD-Calc is a registered trademark of KJ Innovation (see: http://software.kjinnovation.com/GD-Calc.html).

[11] Li, L. et al., Appl. Opt. 38 (1999) 304.

**Figure captions**

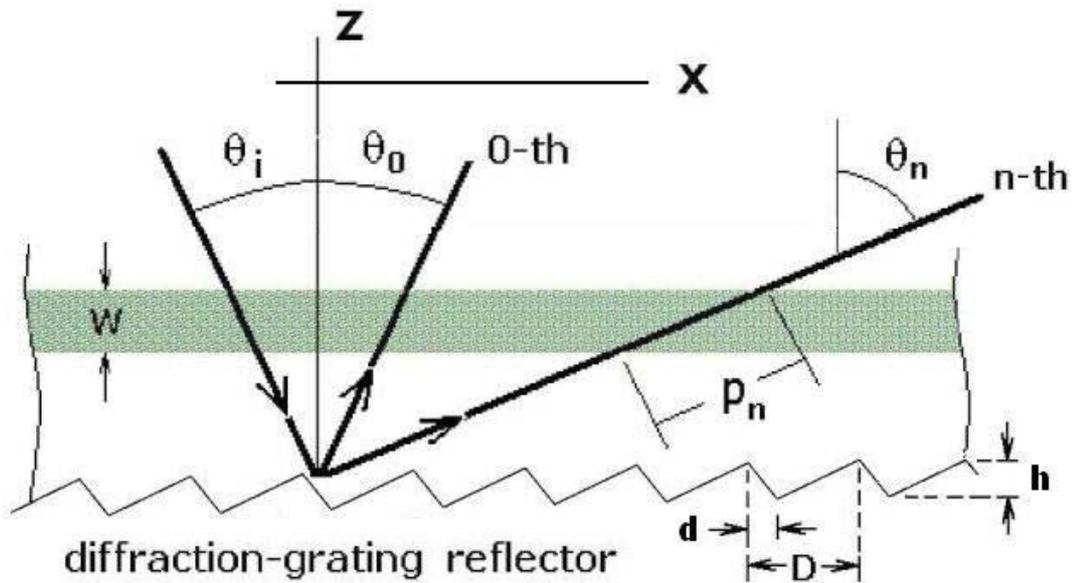

**Figure 1**: Cross section of a PV cell with a diffraction grating reflector with a groove spacing $D$ and an active layer of thickness $W$ (greyed-strip). The incident light splits into several diffraction orders. The diffracted ray of $n$-th order has a path-length $p_n$ within the active zone.

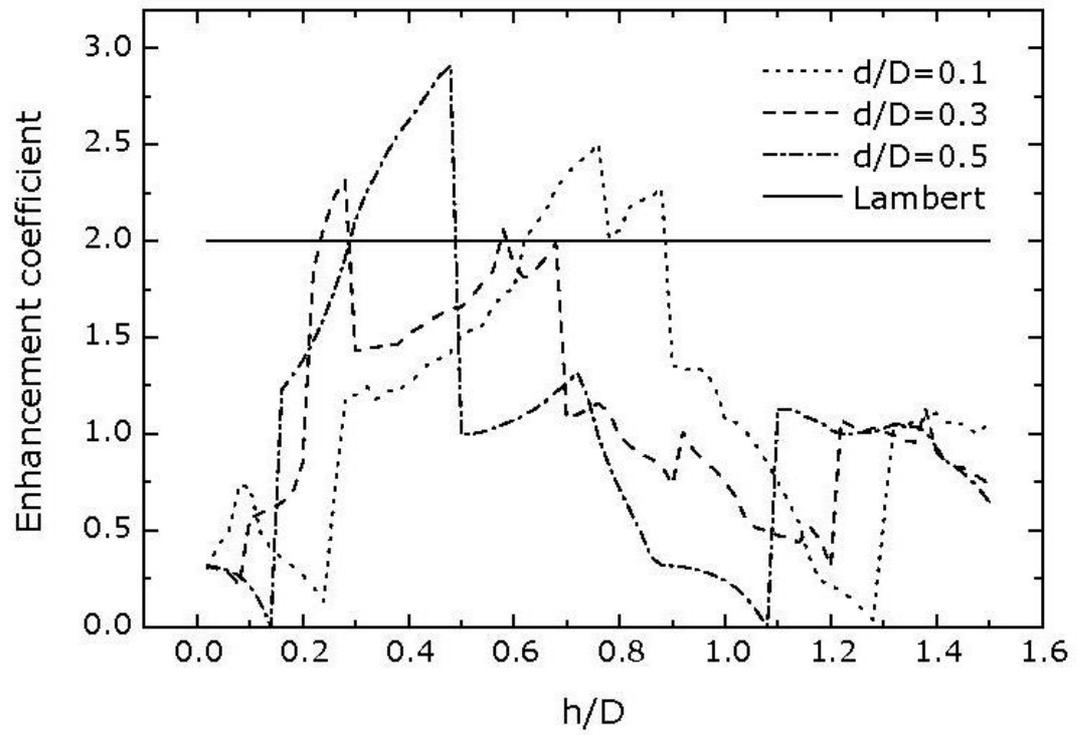

**Figure 2.** Ray-tracing calculated enhancement-coefficient [see Eqs.(3-4)] is shown as a function of the *h*/*D* aspect ratio for TM-polarized light and several *d*/*D*-ratios (see legend). For a comparison, the enhancement expected from a Lambertian reflector with unit reflectivity is denoted as a continuous line

3Preprint submitted to Solar Energy Materials and Solar Cells, November 19, 2008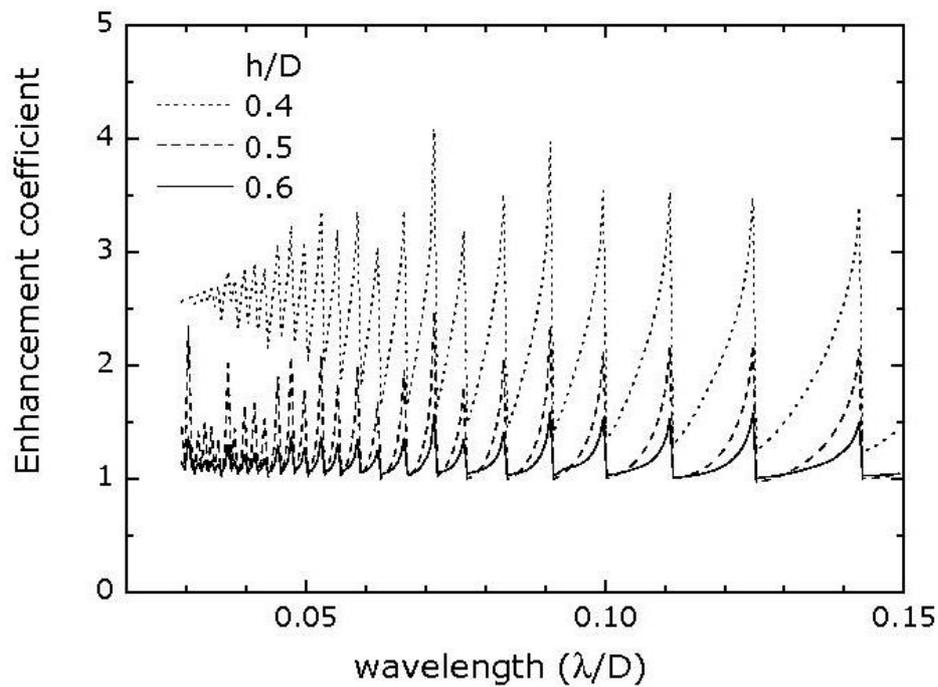

**Figure 3**. Wave-optics calculation of the enhancement-coefficient for normal incidence beams, a diffraction grating with *d/D* = 0.5, and aspect ratios *h/D* = 0.6 (continuous), 0.5 (dashed) and 0.4 (dotted).

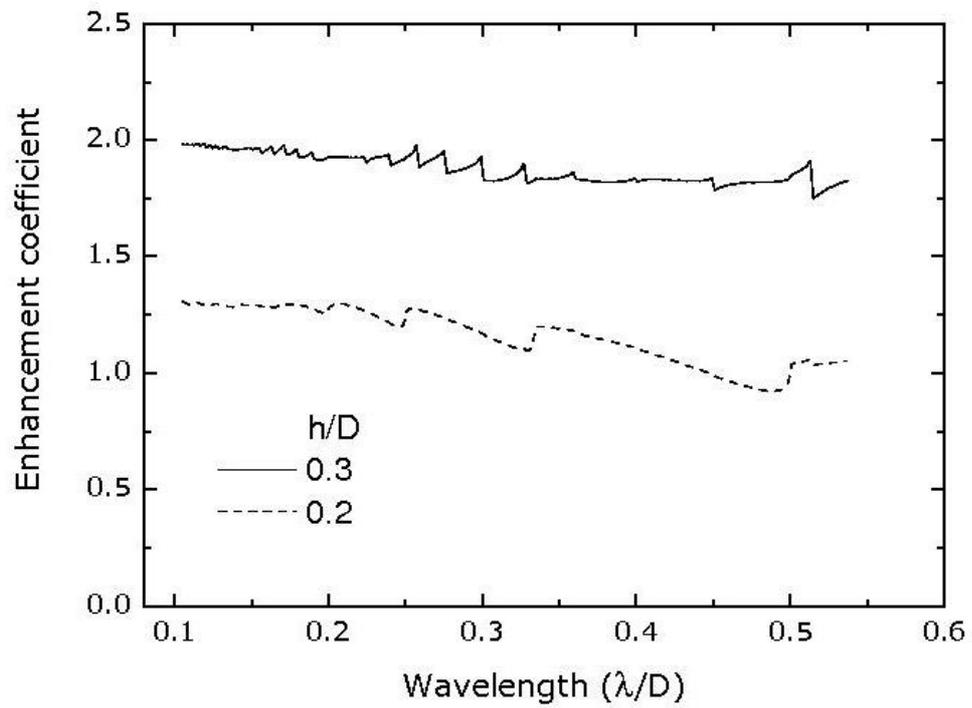

**Figure 4**: Enhancement-coefficient calculated for the same conditions as those of Fig.3 and aspect ratios $h/D$ = 0.3 (continuous) and 0.2 (dashed).